\documentclass[12pt]{article}

\usepackage{latexsym,amsmath,amssymb,theorem,epsfig,graphicx,hyperref,color}

\usepackage{graphicx}
\usepackage{graphicx,subfigure}
\usepackage{epstopdf}
\usepackage{psfrag}
\usepackage{cancel}
 \allowdisplaybreaks[4]

\usepackage[all]{xy}

\DeclareFontFamily{OT1}{rsfs10}{}
\DeclareFontShape{OT1}{rsfs10}{m}{n}{ <-> rsfs10 }{}
\DeclareMathAlphabet{\mathscript}{OT1}{rsfs10}{m}{n}

\numberwithin{equation}{section}

\newcommand{\ns}{\normalsize}

\def\gsim{ \lower .75ex \hbox{$\sim$} \llap{\raise .27ex \hbox{$>$}} }
\def\lsim{ \lower .75ex \hbox{$\sim$} \llap{\raise .27ex \hbox{$<$}} }
\def\be{\begin{equation}}
\def\ee{\end{equation}}
\def\bea{\begin{eqnarray}}
\def\eea{\end{eqnarray}}

\newcommand{\ba}{\begin{array}}
\newcommand{\ea}{\end{array}}

\newcommand{\commentout}[1]{}

\newcommand{\lagm}{{\cal{L}}_m}
\newcommand{\lage}{{\cal{L}}_E}
\newcommand{\detgt}{\sqrt{\tilde{g}}}
\newcommand{\detg}{\sqrt{g}}
\newcommand{\detgg}{\sqrt{\frac{\tilde{g}}{g}}}
\newcommand{\nabt}{\tilde{\nabla}}
\newcommand{\pa}{\partial}
\newcommand{\tmnt}{\tilde{T}^{\mu\nu}}
\newcommand{\tmn}{T^{\mu\nu}}
\newcommand{\mett}{\tilde{g}_{\mu\nu}}
\newcommand{\met}{g_{\mu\nu}}
\newcommand{\cS}{{\cal{S}}}

\newcommand{\comment}[1]{}
\begin{document}

\begin{titlepage}

\title{
  \hfill{\ns }  \\[1em]
   {\LARGE Derivative Chameleons}
\\[1em] }
\author{
     Johannes Noller\footnote{Electronic address: johannes.noller08@imperial.ac.uk}
     \\[0.5em]
     {\ns Theoretical Physics, Blackett Laboratory, Imperial College, London, SW7 2BZ, UK}}

\date{}

\maketitle

\begin{abstract}
We consider generalized chameleon models where the conformal coupling between matter and gravitational geometries is not only a function of the chameleon field $\phi$, but also of its derivatives via higher order co-ordinate invariants (such as $\partial_\mu \phi \partial^\mu \phi, \Box \phi,...$). Specifically we consider the first such non-trivial conformal factor $A(\phi,\partial_\mu \phi \partial^\mu \phi)$. 
The associated phenomenology is investigated and we show that such theories have a new generic mass-altering mechanism, potentially assisting the generation of a sufficiently large chameleon mass in dense environments. The most general effective potential is derived for such derivative chameleon setups and explicit examples are given. Interestingly this points us to the existence of a purely derivative chameleon protected by a shift symmetry for $\phi \to \phi + c$. 
We also discuss potential ghost-like instabilities associated with mass-lifting mechanisms and find another, mass-lowering and instability-free, branch of solutions. This suggests that, barring fine-tuning, stable derivative models are in fact typically anti-chameleons that suppress the field's mass in dense environments. 
Furthermore we investigate modifications to the thin-shell regime and prove a no-go theorem for chameleon effects in non-conformal geometries of the disformal type.
\end{abstract}

\thispagestyle{empty}

\end{titlepage}

\setcounter{tocdepth}{2}
\tableofcontents

\section{Introduction}

Modified gravity theories (see e.g.~\cite{MGreview1,MGreview2} for reviews) have enjoyed continued interest over the past few decades and recent developments have led to a much better understanding of scalar-tensor theories in particular. The notion of screening mechanisms - how a light scalar degree of freedom $\phi$ can act as dark energy on cosmological scales while being shielded in dense environments such as on earth - has turned out to be especially useful in this context. Implementations of such a mechanism include the following: 1) The chameleon model~\cite{KW1,KW2}, where a density dependent mass is generated and the field $\phi$ becomes too massive for detection in dense environments. 2) Vainshtein screened setups~\cite{vainshtein,ags,ddgv} such as DGP~\cite{DGP} and Galileon~\cite{galileon}/Horndeski~\cite{horndeski} models, where non-linear interactions of $\phi$ lead to strongly coupled dynamics. A density-dependent (classical) renormalization of the kinetic energy there results in an effectively decoupled scalar in dense environments. 3) Symmetron models~\cite{symmetron1,symmetron2}, where a scalar $\phi$ is coupled to matter with a coupling strength proportional to the vacuum expectation value of $\phi$. This in turn depends on the ambient density, so that the scalar effectively decouples in high-density regions.  
All these mechanisms reconcile the existence of a light cosmological scalar with tight fifth force constraints on solar-system scales~\cite{solarreview}.

In this paper we wish to focus on chameleon models and potential extensions thereof. Chameleon phenomenology has already been studied extensively~\cite{Gubser:2004uf,Mota:2006fz,Brax:2007ak,Brax:2008hh,Burrage:2008ii,Brax:2009aw,Davis:2009vk,Brax:2009ey,Brax:2010xx,Gannouji:2010fc,Brax:2010uq,Li:2011uw,Mak:2011bt,Brax:2012gr} for variants of the model introduced by~\cite{KW1,KW2}. Typically such theories are built by universally coupling matter to a metric conformally related to the Einstein metric $\met$. Making this bimetric structure explicit we here investigate whether there are any interesting and qualitatively new implementations of the chameleon mechanism that arise when going beyond the simple conformal relationships considered so far. Can the chameleon effect occur in new guises for generic 4D effective field theories?
In other words, we will be asking two questions: What forms can the conformal relation generically take? And does it have to be conformal or are there more general bimetric structures that can produce chameleonic phenomenology? 

A systematic way of undertaking this investigation is to construct a generic matter metric as a function of the Einstein metric $\met$, $\phi$ and derivatives of the field $\pa^n \phi$. By showing how what we dub {\it derivative chameleons} arise in this framework, this paper aims to illustrate how such an approach unveils qualitatively new constructions.    
More specifically we show that derivative chameleons naturally give rise to a new mass-altering mechanism, which changes the mass of oscillations around an effective potential minimum. The mass-lifting branch may help in ensuring $\phi$ can escape detection in fifth force experiments and consequently may be of use in alleviating fine-tuning constraints for chameleon models. However, we also show that care needs to be taken in order to avoid ghost-like instabilities for mass-lifting solutions. The mass-lowering branch of solutions, on the other hand, is generically stable.
Importantly the new mechanism works for purely derivative conformal factors too, opening up the exciting possibility of having a chameleon mechanism which comes endowed with a shift-symmetry in the field $\phi \to \phi + c$, offering better protection from quantum-corrections. 
Furthermore we discuss modifications to the radial solutions around spherical matter sources and modifications to the thin-shell mechanism. 
We also point out why non-conformal geometries are not expected to display chameleon-screening, establishing a no-go theorem for so-called disformal geometries. 

The plan for the paper is as follows. In section \ref{secminimal} we review the standard chameleon picture and how this gives rise to an effective potential with a large mass for oscillations around the effective minimum. Section \ref{secdis} then introduces the bimetric framework, explicitly formulating metric relations which are used to construct chameleon models thereafter.  This construction takes place in \ref{concham2}, while focusing on conformally related metrics, pointing out the phenomenology of our generalized chameleon model, in particular the new mass-altering mechanism. In \ref{secconcrete} this is followed up by presenting a few simple concrete examples that implement this new-found mechanism and show how it can be realized for purely derivative conformal factors. Modifications to thin-shell screening are then discussed while investigating radial solutions around a massive source in section~\ref{secradial}.
In \ref{secdisno} we go beyond conformal relationships, arguing that their natural extension - disformally related geometries - cannot produce chameleon phenomenology (they do, however, naturally generate Vainshtein-type screening solutions). Finally we conclude in \ref{conc} .

\section{Conformal chameleons I: The minimal theory} \label{secminimal}

Chameleon models~\cite{KW1,KW2} are typically constructed with the use of two conformally related metrics, where the conformal factor is a function of the chameleon field $\phi$ only. In particular we may consider an action of the following form
\be \label{Sminimal}
{\cal{S}} = \int d^4 x \detg \left( \frac{M^2}{2}R + X - V(\phi)\right) +  {\cal S}_m\left( A^2(\phi)\met, \Psi_i \right),
\ee
where $X = -\frac{1}{2}\nabla_\mu \phi \nabla^{\mu} \phi$ is the usual canonical kinetic term for $\phi$, $V(\phi)$ is some arbitrary potential and ${\cal S}_m$ describes the matter part of the action with all matter species $\Psi_i$ universally coupled to the matter metric $\mett = A^2(\phi) \met$. This universal coupling ensures the validity of the weak equivalence principle by design. Also note that the signature of $\met$ used here is $(-+++)$, and hence $2X = \dot{\phi}^2 - \left(\vec{\nabla} \phi\right)^2$. 
As usual a matter stress-energy tensor can be defined with respect to the matter (``Jordan frame'') metric $\mett$, so that
\be \label{tmnt}
\tmnt = \frac{2}{\detgt}\frac{\delta \lagm}{\delta \mett}.
\ee
Since it minimally couples to $\mett$, it is covariantly conserved with respect to that metric $\nabt_\mu \tmnt = 0$. When mapping $\tmnt$ into the Einstein frame one finds
\be \label{tmn}
\tmnt = A^{-6}(\phi) T^{\mu\nu}.
\ee
This means $\tmn$ is not covariantly conserved in the Einstein frame $\nabla_\mu \tmn \ne 0$. Turning our attention back to the original action~\eqref{Sminimal}, one can now work out the equation of motion for the scalar $\phi$ obtaining
\be \label{simpleeom}
\Box{\phi} = V_{,\phi} - A^3(\phi) A_{,\phi} \tilde{T},
\ee
where the matter stress-energy tensor is defined as in~\eqref{tmnt} and has been contracted with the matter metric $\tilde{T} = \tmnt \mett$. 
Specializing to the case of a pressureless, non-relativistic source, the only non-vanishing component of the stress-energy tensor $ \tilde{T}^{\mu}_{\nu} $ is $ \tilde{T}^{0}_{0} =-\tilde{\rho}$. Following from~\eqref{tmn} the energy density of matter in the Einstein frame, $\rho$, is given by $\rho =A^4 \tilde{\rho}$. As a direct consequence of $\nabt_\mu \tmnt = 0$ we then also find a conserved quantity in the Einstein frame $\hat \rho = A^3 \tilde \rho = A^{-1} \rho$. In terms of this conserved quantity~\eqref{simpleeom} can therefore be written as
\be \label{potdiv}
\Box{\phi} = V_{,\phi} + A_{,\phi}\hat \rho,
\ee
and one can integrate up to obtain an effective potential for $\phi$
\be \label{standardV}
V_{\rm eff}(\phi) = V(\phi) + \hat \rho A(\phi),
\ee
since $\hat \rho$ is conserved and independent of $\phi$ in the Einstein frame.

\begin{figure}[h] 
\begin{center}$
\begin{array}{cc}
\includegraphics[width=0.45\linewidth]{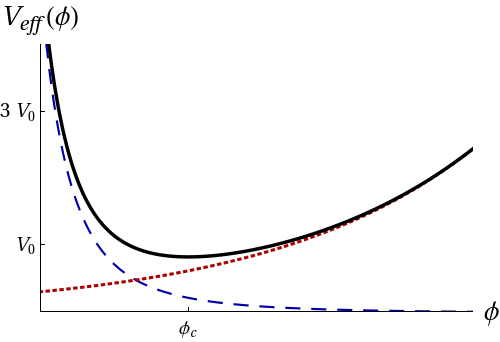} &
\includegraphics[width=0.45\linewidth]{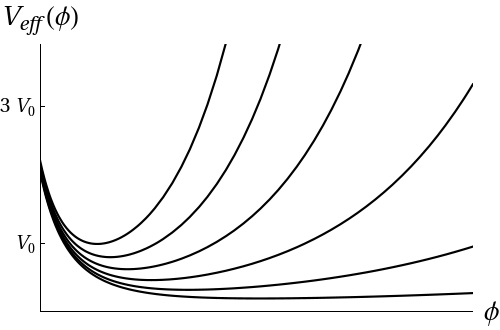}
\end{array}$
\end{center}
\caption{\footnotesize{{\bf Left}: Plot showing the effective potential $V_{eff}(\phi)$ (solid line), the conformal factor $A(\phi) = e^{k_1 \phi}$ (dotted line) and a runaway potential $V(\phi) \sim \phi^{-3}$ (dashed line) in arbitrary units, cf. equation~\eqref{standardV}. {\bf Right}: Plot showing the effective potential for $k_1 = 1-6$ from bottom to top. Note how the position of the minimum $\phi_{min}$ is shifted to smaller $\phi$  and the curvature and hence mass of oscillations around $\phi_{min}$ is enhanced as $k_1$ increases.}}
\label{figconcham}
\end{figure}

This setup has interesting phenomenological consequences. Suppose we start with a runaway potential $V(\phi)$, e.g. $V(\phi) = {M_{Pl}^{n+4}}/{\phi^n}$ as desirable from the point of view of quintessence models~\cite{quintessence}. This ensures that, in the limit when we can ignore the matter action ${\cal S}_m$ (a low-density environment with $\hat \rho \to 0$ in the language set out above), we recover a quintessence-like solution to the cosmological constant problem. A slow-rolling light scalar field then drives accelerated expansion of space-time. 
But in regions of high density this behavior changes in the following way. Naively $\phi$ does not possess a mass term at all, since $V(\phi)$ has no minimum (except for the trivial one at $\phi \to \infty$). However, from~\eqref{potdiv} it becomes clear that, for non-zero $\hat \rho$, a suitably chosen $A(\phi)$ can result in a $V_{eff}$ which does have a minimum $\phi_{\rm min}$, such that 
\be
V_{eff,\phi}(\phi_{\min}) = V_{,\phi}(\phi_{min}) + A_{,\phi}(\phi_{min})\hat \rho = 0,
\ee 
where the mass of the field $m$ for small oscillations around the minimum $\phi_{min}$ is given by
\be
m^2 \equiv V_{eff,\phi\phi}(\phi_{\rm min}) = V_{,\phi\phi}(\phi_{\rm min}) + \hat \rho A_{,\phi\phi}(\phi_{\rm min}).
\ee
This is the essence of the chameleon mechanism: An environmentally-dependent way of generating a large mass for an otherwise very light scalar $\phi$. This reconciles a model such as~\eqref{Sminimal} with fifth force constraints, since $\phi$ becomes too heavy for detection in laboratory experiments on earth, yet can act as dark energy on large scales.
Figure~\ref{figconcham} illustrates the chameleon mechanism for a conformal factor of the form $A(\phi) \sim e^{k_1 \phi}$. 

Two final comments are in order here. Firstly our argument has been purely phenomenological in that a suitably chosen conformal factor can give rise to the chameleon mechanism described, but no arguments for why such an $A(\phi)$ is (technically) natural have been given. This issue is beyond the scope of this paper, but we refer to~\cite{UVcham} for a discussion. 
Secondly we have not discussed the thin-shell mechanism so far, which is another essential ingredient for the viability of the chameleon mechanism in that it suppresses fifth force modifications to e.g. planetary orbits. We will do so in section~\ref{secradial} where radial solutions are investigated in some detail.

\section{The disformal/bimetric perspective}\label{secdis}

The chameleon mechanism presented in the previous section hinges on the existence of two conformally related metrics $\met$ and $\mett$, respectively used to construct the gravitational part of the action (the Ricci scalar) and the matter part of the action (i.e. matter fields $\Psi_i$ are minimally coupled to $\mett$). The question we want to answer in this paper is: What happens when this relationship is modified? To be more specific, are there other classes of scalar-tensor theories that offer qualitatively distinct implementations of the chameleon mechanism (which we take to be an environmentally dependent generation of mass for an otherwise light cosmological scalar $\phi$)?

\subsection{A bimetric scalar-tensor theory} \label{bimst}

In this section we want to lay out the problem in more general terms without imposing as restrictive a metric relation as in section~\ref{secminimal}. Consequently let us start with a schematic action of the following form
\be \label{Sgeneral}
{\cal{S}} = \int d^4 x \detg \frac{M^2}{2}R + {\cal S}_m\left( \mett, \Psi_i \right) + \cS_\phi
\ee 
where $\Psi_i$ are matter fields minimally coupled to $\mett$ as before and $\cS_{\phi}$ denotes an action giving the scalar field $\phi$ dynamics of its own. We emphasize that there is no a priori requirement constraining $S_\phi$ to be formed with either $\met$ or $\mett$. To enable comparison with the existing chameleon literature we fix the form of $\cS_\phi$ such that
\be \label{chamact}
{\cal{S}} = \int d^4 x \detg \left( \frac{M^2}{2}R + X - V(\phi)\right) +  {\cal S}_m\left( \mett, \Psi_i \right),
\ee
i.e. $\cS_\phi$ equips $\phi$ with a canonical kinetic term and a potential minimally coupled to $\met$. In order to investigate~\eqref{chamact} it now becomes necessary to specify how $\met$ and $\mett$ are related, and in particular how $\phi$ enters this relation. For this it will be useful to schematically write~\eqref{chamact} as 
\be
{\cal S} = \int d^4 x \detg {\cal L}_E \left( \met, \phi \right) + \int d^4 x \detgt {\cal L}_m \left( \mett, \Psi_i \right).
\ee

\subsection{The metric relation}

In section~\ref{secminimal} gravity and matter metrics were conformally related by
\be
\mett = A^2(\phi) \met.
\ee 
Here we are interested in investigating more general metric relations. A systematic way of constructing a more general matter metric in the presence of a gravitational scalar $\phi$ is to write
\be
\mett = \mett(\met,\phi,\pa\phi,\pa\pa\phi,...),
\ee
which is therefore a function of the gravity metric $\met$, the field $\phi$ and its derivatives.\footnote{While this paper was being finished an analogous construction for a metric that is a composite of several other fields $\phi_i$ was investigated in~\cite{kimpton}.} We will truncate this expansion in derivatives at first order, i.e. $\Box \phi$ contributions and other higher order terms are ignored. If~\eqref{chamact} is an effective field theory then subsequently higher derivatives of the field can be suppressed by higher powers of the cutoff scale, motivating such a truncation. 
Note, however, that such an argument naturally does not discriminate between e.g. $(\pa \phi)^n$ and $\pa^n \phi$. So in addition, and more heuristically perhaps, we will allow operators such as $(\pa \phi)^n$, while disregarding $\pa^n \phi$ for $n = 2$ in particular. This setup trivially protects us from having equations of motion with higher than second order derivatives of the field, where we would expect ghost-like instabilities via Ostrogradski's theorem~\cite{Ostro}.    

Arguing in a similar fashion, Bekenstein postulated~\cite{bekenstein} that the most general gravitational and matter metrics satisfying these conditions\footnote{In fact Bekenstein assumes what is essentially a 4D effective field theory in an overall Finslerian geometry with gravity and matter metrics related by a single degree of freedom $\phi$ up to first order in its derivatives. He then proceeds to show that this has to reduce to a Riemannian geometry described by \eqref{bim}. For details see~\cite{bekenstein}.} give rise to a bimetric theory where $\mett$ and $\met$ are ``disformally'' related by
\be
\mett = A^2(\phi,X)\met + B^2(\phi,X)\pa_\mu \phi \pa_\nu \phi, \label{bim}
\ee
where we again recognize $X$ as $\phi$'s kinetic term. 
Note that we have implicitly required the metric relation to only be a function of coordinate invariants (hence the dependence of $A^2$ and $B^2$ on $X$ and not $\partial_\mu \phi$ by itself).
This generalized metric relation therefore consists of a conformal piece, with $A^2\left(\phi,X\right)$ being the conformal factor, and a so-called disformal piece, where we shall call $B^2\left(\phi,X\right)$ the disformal factor. Interestingly the disformal piece of the metric relation exactly mimics the structure of induced metrics in models that are intrinsically higher-dimensional in nature. For example, in higher-dimensional dark energy setups such as DGP~\cite{DGP} or DBI galileons~\cite{DBIgalileon} the induced metric on our 4-brane is $\mett = \met + \pa_\mu \phi \pa_\nu \phi$, i.e. exactly of the disformal type with a trivial disformal factor $B^2 = 1$.

\subsection{Generalized equations of motion}

With a matter stress-energy tensor defined as in~\eqref{tmnt} and metrics related by~\eqref{bim}, we can now write down the equations of motion for the general action~\eqref{chamact}. Varying~\eqref{chamact} we find
\bea
\frac{\partial \lage}{\pa \phi} + \frac{1}{2}\detgg \tmnt \frac{\pa \mett}{\pa \phi} &=& \nabla_\alpha \frac{\pa \lage}{\pa \phi_{,\alpha}} + \frac{1}{2}\frac{\pa \mett}{\pa \phi_{,\alpha}} \nabla_{\alpha} \left(\detgg\tmnt   \right) + \frac{1}{2}\detgg\tmnt \nabla_\alpha \frac{\pa \mett}{\pa \phi_{,\alpha}}.\label{phieom}
\eea
Note that these expressions neither assume anything about the form of $\tmnt$ nor rely on a conformal relationship between $\met$ and $\mett$ as considered above.
It is, however, worth emphasizing that $\tmnt$ is explicitly a stress energy tensor not including any contributions from $\cS_\phi$ in~\eqref{Sgeneral}, so it does not satisfy the Einstein equations by itself. Instead, after having mapped all quantities into the Einstein frame, one finds $G_{\mu\nu} = 8\pi G \left(T_{\mu\nu}^{(\text{matter})} + T_{\mu\nu}^{(\phi)} \right)$.

\section{Conformal chameleons II: Derivative setups} \label{concham2}

Here we wish to focus on the effect of reducing the bimetric relationship~\eqref{bim} to the simple case $B(\phi,X) = 0$. This is the most general purely conformal relation permitted by~\eqref{bim}. The conformal factor $A^2(\phi,X)$ is an arbitrary function of the field $\phi$ and the first higher order coordinate invariant $X$, $\phi$'s kinetic term, and we have
\be
\mett = A^2(\phi,X)\met. 
\ee
As such this means the full action under consideration is that of a generalized conformal chameleon
\be \label{Shigher}
{\cal{S}} = \int d^4 x \detg \left( \frac{M^2}{2}R + X - V(\phi)\right) +  {\cal S}_m\left( A^2(\phi,X)\met, \Psi_i \right).
\ee
The relation between matter stress-energy tensors in different frames straightforwardly generalizes to $\tmnt = A^{-6}(\phi,X) T^{\mu\nu}$. We will now first compute the associated equation of motion for $\phi$ and then comment on the impact a generalized conformal factor has on the chameleon mechanism.

\subsection{Equation of motion and effective potential}

The equation of motion for $\phi$ that follows from~\eqref{Shigher} can be written in the following form using~\eqref{phieom}
\begin{eqnarray}\label{eom3}
-V_{,\phi} + A_{,\phi} A^3 \tilde{T} &=& - \Box\phi + \sum_i {\cal J}_i, 
\end{eqnarray}
where $\Box \equiv \nabla^\mu \nabla_\mu$ and
\begin{eqnarray}
{\cal J}_1 &=&  - A_{,X} A^3 \tilde{T} \Box \phi \nonumber \\
{\cal J}_2 &=&  2 A^2 \tilde{T} \left( A A_{,X\phi} + 3 A_{,X} A_{,\phi} \right) X \nonumber \\
{\cal J}_3 &=& A^2 \tilde{T} \left( A A_{,XX} + 3 A_{,X}^2 \right) \Pi \nonumber \\
{\cal J}_4 &=& -A^3 A_{,X} \mett \pa^{\alpha} \phi \tilde{\nabla}_\alpha \tmnt,
\end{eqnarray}
and we have defined $\Pi \equiv \nabla_\mu \nabla_\nu \phi \nabla^\mu \phi \nabla^\nu \phi = \nabla_\mu \partial_\nu \phi \partial^\mu \phi \partial^\nu \phi$.
We have assumed that $\tilde{T}^{\mu\nu}$ is symmetric and expressed derivatives acting on $\tmnt$ in terms of matter frame variables to avoid mixing operators and variables defined for different frames. More explicitly this means using
\begin{eqnarray}
\nabla_\alpha \tilde{T}^{\mu\nu} =  \tilde{\nabla}_\alpha \tilde{T}^{\mu\nu} - 2 \Gamma^{(\mu}_{\beta\alpha} \tilde{T}^{\nu)\beta},
\end{eqnarray}
where $\Gamma^{\mu}_{\beta\alpha}$ is the connection associated with transformations between $\mett$ ($\sim$ Jordan) and $\met$ (Einstein) frames (for details please see the appendix). 
Additionally there are two further types of terms one might naively expect to arise from varying~\eqref{Shigher} 
\begin{eqnarray} \label{j56}
{\cal J}_5 &\propto&  A^4 A_{,X}^2 \tilde{T}^{\beta\nu}\Pi_{\nu\rho}\pa_\beta \phi \pa^{\rho} \phi \nonumber \\
{\cal J}_6 &\propto& A^4 A_{,X} A_{,\phi} \pa_\beta \phi \pa_\nu \phi \tilde{T}^{\beta\nu},
\end{eqnarray}
where $\Pi_{\nu\rho} \equiv \nabla_\nu \nabla_\rho \phi = \nabla_\nu \pa_\rho \phi$.
However, the symmetry imposed on the conformal factor (i.e. $A$ only being a function of coordinate invariants) means that contributions proportional to ${\cal J}_5$ and ${\cal J}_6$ cancel and hence do not appear in the equation of motion. As an immediate consequence there is no direct, Vainshtein-like coupling between the stress-energy tensor and derivatives of the field, but matter only enters via $\tilde{T}$.\footnote{Note that, at any rate, the Vainshtein-like coupling referred to here would be a derivative coupling which leads to a screening, i.e. a classical renormalization of the kinetic energy, that depends locally on $\tmn$. This means that there will be no screening effect even a small distance away from the source. This is in contrast to Vainshtein screening sourced by derivative self-interactions of the scalar $\phi$, e.g. a $X\Box \phi$ term in the action, which will lead to screening inside a Vainshtein radius $r_V$ that can extend beyond the source itself, subject to introducing an appropriate coupling between matter and $\phi$ such as the linear $\phi \tilde{T}$.}  Taking $\mett$ inside its covariant derivative and using that $\tilde{\nabla} s = \nabla s = \pa s$ for any scalar $s$ (i.e. covariant derivatives related to both metrics act on scalars in the same way), the overall equation of motion can therefore be written
\begin{eqnarray} \label{fulleom}
-V_{,\phi} + A_{,\phi}A^3 \tilde{T} = &-& \left( 1 + A^3 \tilde{T} A_{,X} \right)  \Box \phi \nonumber \\
&+& 2 A^2 \tilde{T} \left( A A_{,X\phi} + 3 A_{,X} A_{,\phi} \right) X \nonumber \\
&+& A^2 \tilde{T} \left( A A_{,XX} + 3 A_{,X}^2 \right) \Pi \nonumber \\
&-& A^3 A_{,X} \pa^{\alpha} \phi \pa_\alpha \tilde{T}.
\end{eqnarray}

We now proceed to simplify these expressions by considering a uniform matter source, making the following key assumption about the system under consideration
\begin{itemize}
\item The stress-energy tensor describes a pressureless, non-relativistic fluid ($\tilde{T}_0^0 = -\tilde{\rho}$) with all other stress-energy tensor components vanishing.
\end{itemize}
Later on (in section~\ref{secradial}) we will also assume a static, uniform source in the matter frame ($\tilde{\nabla}_\alpha \tilde{T}^{\mu\nu} = 0$ inside the source). Combining this with the first assumption, this is equivalent to assuming $\tilde{\nabla}_\alpha \tilde{\rho} = \pa_\alpha \tilde{\rho} = 0$ (again, inside the source). This means ${\cal J}_4 = 0$ everywhere except in the transition between source and surroundings. For this reason we will keep the full ${\cal J}_4$ term when evaluating radial profiles across this boundary and not make any further assumptions about ${\cal J}_4$ for the time being. Note that we also do {\it not} assume a static profile for $\phi$ ($\partial_0 \phi = 0$), which one may want to impose for further simplification in a late-universe system like the solar system, which has had time to settle. Modeling matter as a pressureless, non-relativistic fluid results in 
\begin{eqnarray} \label{Jequations}
{\cal J}_1 &=&  A_{,X} \hat\rho \Box \phi \nonumber \\
{\cal J}_2 &=& - 2 \hat\rho \left( A_{,X\phi} + 3 A^{-1} A_{,X} A_{,\phi} \right) X \nonumber \\
{\cal J}_3 &=& - \hat\rho \left( A_{,XX} + 3 A^{-1} A_{,X}^2 \right) \Pi \nonumber \\
{\cal J}_4 &=&  A^3 A_{,X} \pa^{\alpha} \phi {\pa}_\alpha \left( A^{-3} \hat\rho \right),
\end{eqnarray}
where we have substituted $A^3 \tilde{T} = - \hat{\rho}$ (or equivalently $\hat \rho = A^3 \tilde{\rho}$), since $\hat{\rho}$ is a conserved quantity in the Einstein frame. Explicitly working out ${\cal J}_4$, which will be relevant when modeling the transition across matter boundaries, one finds
\begin{eqnarray}
{\cal J}_4 &=&   6 A^{-1} A_{,X} A_{,\phi} \hat\rho X + 3 A^{-1} A_{,X}^2 \hat \rho \Pi + A_{,X} \pa^{\alpha} \phi \pa_\alpha \hat\rho.
\end{eqnarray}
This considerably simplifies the equations of motion, revealing additional symmetries that arise due to the functional form of $A$ in combination with requiring a non-relativistic, pressureless fluid as the matter source. Consequently the equation of motion for $\phi$ may be written as
\begin{eqnarray} \label{theeom}
V_{,\phi} + A_{,\phi}\hat\rho &=& \left( 1 -\hat\rho A_{,X} \right)  \Box \phi + 2 \hat\rho A_{,X\phi} X + \hat\rho A_{,XX} \Pi - A_{,X} \pa^{\alpha} \phi \pa_\alpha \hat\rho. 
\end{eqnarray}

An instructive way to think of the physical properties of this system is to explicitly write it as a Klein-Gordon equation with an effective potential that only depends on $\phi$ and ``friction terms'' that encode the dependence on higher derivatives of $\phi$. If we are purely interested in the profile inside the source (and hence ignore $\pa_\alpha \hat\rho$), this means we can write  
\begin{eqnarray} \label{eom4}
V_{eff,\phi}(\phi) &=& \Box\phi + {\cal F}_1\left(\phi,X,\hat\rho\right) X + {\cal F}_2\left(\phi,X,\hat\rho\right) \Pi \nonumber \\
&=& \Box \phi + \text{``friction terms''}.
\end{eqnarray}
To make the form of the effective potential explicit, one can Taylor-expand the conformal factor $A$ in powers of $X$, writing
\be \label{Taylor}
A(\phi,X) = A^{(0)}(\phi) + A^{(1)}(\phi)X + {\cal O}(X^2),
\ee
which allows us to write down the effective potential for $\phi$ as
\be \label{Vprime}
V_{eff,\phi}(\phi) = \left(V_{,\phi} + \hat\rho A^{(0)}_{,\phi}\right) \left( 1 - A^{(1)} \hat \rho \right)^{-1}.
\ee

\subsection{Phenomenology and comments} 

Given some original potential of the runaway form $V(\phi)$, the chameleon mechanism generates an environmentally dependent effective potential that gives $\phi$ a large mass in high density regions. Here ``high density'' and ``large mass'' are essentially references to solar system constraints~\cite{solarreview} on the presence of a fifth force mediated by a scalar degree of freedom. As such, any theory with a runaway $V(\phi)$ that successfully implements the chameleon mechanism at the very least has to tick two boxes. Firstly it needs to give rise to an environmentally-dependent effective potential which has a minimum. And secondly the mass of small oscillations around that minimum has to be large enough to satisfy fifth force constraints. With this in mind let us investigate the effective potential described by~\eqref{theeom} and~\eqref{Vprime}. Note that in section~\ref{secconcrete} we will give explicit examples illustrating each of the effects outlined here in detail.

{\bf Position of minimum:} 
For a finite density $\hat \rho$ and conformal factor $A$ (more precisely, $A^{(1)}$),  \eqref{Vprime} shows that a potential minimum requires
\be
V_{,\phi}\left(\phi_{\min}\right) + \hat\rho A^{(0)}_{,\phi} \left(\phi_{min}\right) = 0,
\ee
which is identical to the condition for the minimal conformal chameleon discussed in section~\ref{secminimal}. In other words, the position of the minimum, $\phi_{min}$, is not altered by the introduction of an $X$-dependent conformal factor. This also shows that a conformal factor which does not depend on $\phi$ itself, but only on derivatives of $\phi$, (i.e. $A^{(0)}_{,\phi} = 0$, as is the case for $A = A(X)$) cannot generate an effective potential with a minimum, if the original $V(\phi)$ does not already possess a minimum itself. The $\phi$-dependence of $A$ is consequently essential to obtaining a successful implementation of the chameleon mechanism, if starting with a runaway potential (e.g. $V(\phi) \sim \phi^{-n}$). A potential $V(\phi)$ with very small mass $m^2_V$ can however be uplifted by a pure derivative conformal factor (e.g. $A(X)$), leading to an effective potential with mass $m^2 \gg m^2_V$. Below and in the next section we will give explicit examples illustrating this behavior.

{\bf Effective mass:} 
A derivative-dependent conformal factor affects the curvature of the effective potential. This means derivative chameleons generically come equipped with a new mass-altering mechanism. The effective potential, and hence the mass of the field, is classically renormalized by introducing higher order invariants into the conformal relation as follows
\begin{eqnarray}  \label{mass}
m^2 = V_{eff,\phi\phi}(\phi_{min}) &=& \left(V_{,\phi\phi}(\phi_{min}) + \hat\rho A^{(0)}_{,\phi\phi}(\phi_{min})\right) \left( 1 - A^{(1)}(\phi_{min}) \hat \rho \right)^{-1} \nonumber \\
&=& m^2_{standard} \left( 1 - A^{(1)}(\phi_{min}) \hat \rho \right)^{-1},
\end{eqnarray}
where $m^2_{standard}$ denotes the effective mass for small oscillations around the minimum for a theory with identical conformal factor in the limit $A(\phi,X \to 0)$. In other words, $m^2_{standard}$ is the effective mass for the theory in the limit where higher-derivative contributions can be neglected. This mass-altering mechanism can be separated into three branches, which we will discuss now.  

{\bf Mass-lifting, ghost-like instabilities and anti-chameleons}:
Equation~\eqref{mass} shows that the effective mass $m^2$ is enlarged when  $0 < A^{(1)}(\phi_{min})\hat\rho < 1$. 
This is interesting since it suggests that derivative chameleon models provide an additional mass-lifting mechanism, potentially alleviating the fine-tuning involved in obtaining a sufficiently large mass in dense environments for standard chameleon models. 
However, care must be taken when considering mass-lifting solutions for the following reason. Suppose we consider a conformal factor $A$ such that a mass-lifting mechanism is in place, i.e. $0 < A^{(1)}(\phi_{min})\hat\rho_1 < 1$ for some given energy density $\hat\rho_1$.  Now, assuming $A$ has no density-dependence itself, one can solve for a (larger) critical density $\hat \rho_{crit}$ above which the solution becomes unstable. The effective potential switches sign since $1 - A^{(1)}(\phi_{min})$ becomes negative, so that $\phi_{min}$ becomes $\phi_{max}$ and $V_{eff,\phi\phi}(\phi_{min})$ turns negative. Thus we are left with a negative ``mass term'', signaling instabilities, and the solution becomes ghost-like. Figure~\ref{figexample} illustrates these different regimes. That different energy densities $\hat\rho$ will interpolate between stable mass-lifting and ghost-like solutions can also be seen from the relevant part of the equation of motion 
\begin{eqnarray}
V_{,\phi} + A_{,\phi}\hat\rho &=& \left( 1 -\hat\rho A_{,X} \right)  \Box \phi + ...,
\end{eqnarray}
where, if $\hat\rho A_{,X} > 1$, this can be traced back to an action with the ``wrong'' sign for $\phi$'s kinetic term.\footnote{Jumping ahead slightly, also note that, if $A_{,X}$ is a function of $\phi$ and the field is not approximately constant $\phi \sim \phi_{min}$ inside the source (the so-called ``thick-shell regime - see section~\ref{secradial}), one needs to be aware that $\hat\rho A_{,X}(\phi)$ being smaller than unity for some initial $\phi_i (r = 0)$ no longer guarantees that this remains true for all values of $\phi$ taken inside the source.} 

There appear to be two obvious solutions to this instability problem.\footnote{Should we decide to bite the bullet and accept the existence of ghost-like solutions above some density-scale $\hat\rho$, such an approach will also face major challenges when confronted with high energy/density  early universe physics.} Firstly one could consider making $A(\phi,X)$ a function of the energy density $\hat\rho$ as well. However, especially given that the matter stress-energy tensor $\tilde{T}^{\mu\nu}$ {\it is} a variation of the matter Lagrangian with respect to the metric $\mett = A^2(\phi,X)\met$, it is not clear what such an iterative dependence on $\tilde{T}^{\mu\nu}$ would mean. Nevertheless it will be an interesting task for the future to think about whether there is some convincing way of implementing such a dependence. In any case, from a purely phenomenological point of view, a density dependent $A$ allows us to have a stable, derivative dependent mass-lifting mechanism for all $\hat\rho$. Secondly one may choose a conformal factor such that $\hat\rho A_{,X} \lsim 1$ up to the density cutoff of the theory.
Note, however, that the effective chameleon mass $m^2$ is proportional to $\left(\hat\rho_{crit} - \hat\rho\right)^{-1}$, where we have defined a critical density $\hat\rho_{crit}$ such that $A^{(1)}(\phi_{min})\hat\rho_{crit} = 1$. Introducing a cutoff at $\hat\rho_{crit}$ will then render derivative-dependent effects on $m^2$ suppressed by that same cutoff scale 
\be
m^2 = \frac{m^2_{standard}}{\left( \hat\rho_{cutoff} - \hat \rho\right) A^{(1)}(\phi_{min})} = m^2_{standard} \left( 1-\frac{\hat\rho}{\hat\rho_{cutoff}}  \right)^{-1}.
\ee
If the cutoff is low enough, significant derivative-dependent effects on $m^2$ may still be obtained, but for a high cutoff density such as the Planck density $\rho_{P}$ they will be strongly suppressed.

The third branch of mass-altering solutions corresponds to the case when $A^{(1)}(\phi_{min}) < 0$? Then the effective mass of $\phi$ is reduced, counter-acting chameleon screening effects 
\be
m^2 = \frac{m^2_{standard}}{ 1 + \left|A^{(1)}(\phi_{min}) \hat \rho \right|}.
\ee
This branch is free of instabilities and provides a robust mechanism to suppress mass terms in models with derivative conformal factors $A$, since the derivative dependence reduces the curvature of $V_{eff}$. This suggests that the simplest $A^2(\phi,X)$ models, where $A$ is independent of $\hat\rho$ and no ghost-like instabilities arise for any $\hat\rho$, are anti-chameleon models in the sense that the effective mass $m^2$ is reduced compared with the non-derivative chameleon limit $A(\phi, X \to 0)$.

{\bf Thin shell regime:}
The computation of $V_{eff}$ and its minima/maxima and masses above was oblivious to ``friction'' terms in~\eqref{eom4} (by definition, since the effective potential is a function of $\phi$ and not its derivatives). But such terms automatically arise as a consequence of a higher order conformal coupling. Here ``friction terms'' is a reference to terms other than $\Box \phi$ that have a derivative dependence on $\phi$, e.g.  ${\cal J}_2$ and ${\cal J}_3$ which encode the dependence on $X$ and $\Pi$. While not influencing $V_{eff}$, these terms do impact the dynamics of $\phi$, warranting further investigation. 

One particularly interesting consequence is a modification of the so-called thin-shell regime. In typical chameleon theories the chameleon-charge of large, massive objects can be Yukawa-screened with only a thin-shell on the outside of the object contributing to the exterior $\phi$-profile~\cite{KW1,KW2}. This is essential for e.g. avoiding unacceptably large effects on planetary orbits due to a chameleonic fifth force. 
Now the presence of additional friction terms modifies the gradient of $\phi$ inside the source and one therefore expects a modification of the thin-shell effect as well. We will discuss this in detail in section~\ref{secradial}, where we investigate radial solutions around massive sources, focusing on new-found phenomenology due to derivative conformal factors.

{\bf Equivalence principle (violations):}
A final comment on possible equivalence principle violations. By construction, chameleon models (both standard as well as the derivative generalization considered here) respect the weak equivalence principle. In the non-derivative case, the extra degree of freedom $\phi$ locally influences the dynamics, but it does not discriminate between different test masses/types of matter due to the universal coupling of $A^2(\phi)$ to all matter fields $\Psi_i$ (field-dependent couplings are discussed in~\cite{fielddep}). In the derivative case, an additional degree of freedom $X$ enters. But since all matter still universally couples to $A^2(\phi, X) \met$, all test masses locally experience the same gravitational force (note that $\phi$ is viewed as a gravitational scalar here). In both cases $\phi$'s profile of course does not remain constant across space-time, as it depends on the ambient density. This means the strong equivalence principle is trivially violated (as is, in fact, the Einstein equivalence principle - cf. with e.g. the Horndeski constructions in~\cite{fabfour} where the strong, but not the Einstein equivalence principle is broken). 

The main point we wish to make here is that a prima facie worry one might have when considering $A^2(\phi,X)$, namely that a derivative-dependent coupling to matter violates even the weak equivalence principle, is not justified. This is simply due to the fact that there is no dependence of the gravitational coupling on the momentum of matter ($\Psi_i$) test masses, but only a derivative coupling to the gravitational scalar $\phi$. 
Nevertheless this is an area that warrants further investigation, since known violations of the strong equivalence principle in chameleon models (cf.~\cite{Adams,Hui,Chang,Davis,Pourhasan:2011sm}) will be modified by introducing a derivative dependence. Computing this in detail should enable further disentangling of derivative vs. non-derivative chameleons.

\section{Effective potentials and the chameleon mass} \label{secconcrete}

In this section we illustrate how chameleon mechanisms arise in derivative theories with a number of explicit examples. We focus on
the mass-altering mechanism for different conformal factors.

\subsection{Example I: Taylor-expanding $A(\phi,X)$} 

We begin by reminding ourselves of~\eqref{Taylor} and Taylor-expanding the conformal factor
\be
A(\phi,X) = A^{(0)}(\phi) + A^{(1)}(\phi) X + {\cal O}(X^2),
\ee
where the associated mass of oscillations around an effective minimum is given by~\eqref{Vprime}. We will ignore higher orders in $X$ and, as a first simple example, focus on the zeroth order contribution in $\phi$ to $A^{(1)}$, treating it as a constant. As such the conformal factor takes on the form
\be \label{simpleTaylor}
A(\phi,X) = A^{(0)}(\phi) + k_X X.
\ee
The resulting potential and effective mass can be computed to give
\begin{eqnarray} \label{kxeom}
V_{eff}(\phi) &=& \frac{V(\phi) + \hat\rho A^{(0)}(\phi)}{1 - k_x \hat\rho}, \\
m^2 &=& \frac{V_{,\phi\phi}(\phi_{min}) + \hat\rho A^{(0)}_{,\phi\phi}(\phi_{min})}{1 - k_X \hat\rho} = \frac{m^2_{standard}}{1-k_X \hat\rho}.
\end{eqnarray}
This illustrates the point made in the previous section about the existence of a mass-raising and a mass-lowering branch for derivative chameleon models. In the simple setup considered here, if $k_X$ is negative, the chameleon mass $m^2$ is reduced compared with the standard $A = A(\phi)$ theory. The fine-tuning necessary in order to get a sufficiently large chameleon mass is therefore made more severe in this case, since an additional mass-lowering mechanism is at work. For positive $k_X$ a mass-lifting mechanism operates, increasing $m^2$. However, for $k_X \hat\rho = 1$ the mass diverges and, once $k_X \hat\rho$ exceeds unity, the ``mass'' turns negative and the solution ghost-like. As discussed in the previous section, this is cause for concern, since given some gravitational theory the conformal factor is fully specified (and hence $k_X$ is fixed in this case). As such, sources with densities above $\hat\rho_{crit} = 1/k_X$ essentially see an unstable inverted potential $-V_{eff}$ inside the source.  The bottom left graph of figure~\ref{figexample} illustrates this point by plotting the dependence of the effective potential on $k_X$ for some given source density $\hat\rho_1$. This amounts to considering different normalizations of the effective potential. On the other hand, the bottom right plot shows how varying $\hat\rho$ for some given, positive and fixed $k_X$ affects the solution. Note that having restricted to positive $k_X$ means there are no mass-lowering solutions present in this plot. As expected from~\eqref{kxeom}, changing $\hat\rho$ modifies both the position of the minimum as well as as changing the potential's normalization and hence curvature/mass via the $1 - k_x \hat\rho$ term.

As an aside, notice that, if one does allow the conformal factor to depend on energy density $\hat\rho$ (for a discussion of this approach see the previous section), then one can construct a solution which remains ghost-free inside sources with arbitrarily large energy densities $\hat\rho$. E.g. we may impose
\be
A(\phi,X) = e^{k_1 \phi} + {\hat\rho}^{-1} \left( 1 - e^{k_2 \phi} \right)  X + {\cal O}(X^2),
\ee
which results in an effective mass given by
\be
m^2 = e^{k_2 \phi_{min}} \left( V_{,\phi\phi}(\phi_{min}) + k_1^2 \hat\rho e^{k_1 \phi_{min}} \right).
\ee
In other words, in this particular example $k_2$ allows tuning the mass $m^2$ arbitrarily. The potential is stable for all $\phi$ and the mass can be altered by modifying the $X$-dependence of the conformal factor $A^2$.\footnote{If one is concerned that additional singularities are introduced by the $\hat\rho^{-1}$ dependence of $A$ here, notice that a condition such as $A^{(1)}\hat\rho \sim 1- e^{-\hat\rho}$ will also ensure a ghost-free mass-lifting regime.} 

\subsection{Example II: A separable $A(\phi,X)$}

\begin{figure}[h] 
\begin{center}$
\begin{array}{cc}
\includegraphics[width=0.45\linewidth]{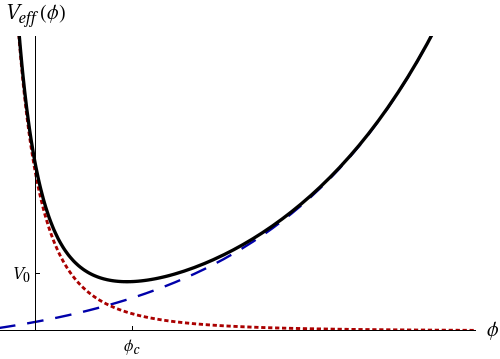} &
\includegraphics[width=0.45\linewidth]{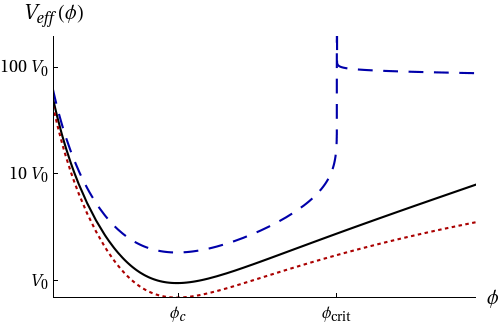} \\
\includegraphics[width=0.45\linewidth]{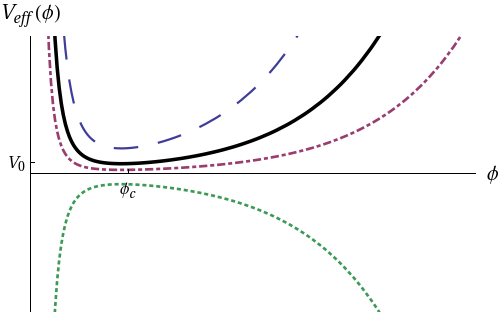} &
\includegraphics[width=0.45\linewidth]{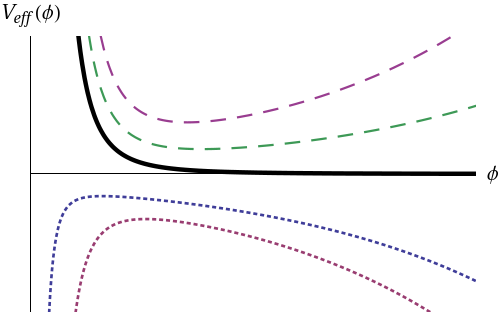}
\end{array}$
\end{center}
\caption{\footnotesize{{\bf Top Left}: Plot showing the effective potential $V_{eff}(\phi)$ (solid line) for a separable conformal factor $A(\phi,X) = B(\phi)C(X) = e^{k_1\phi + k_2 X}$ (dashed line)  and a runaway potential $V(\phi) \sim \phi^{-3}$ (dotted line) in arbitrary units. 
{\bf Top Right}: Logarithmic plot showing $V_{eff}(\phi)$ for different choices of $k_2$ (negative, zero and positive from bottom to top with $\hat\rho = 1$ in arbitrary units). This shows how the amplitude, and hence mass, of the chameleon field grows as $k_2$ is enlarged, while the stable region of parameter space is reduced. The dashed line shows how the pole in the denominator of $V_{eff}$ creates a discontinuity at $\phi_{crit}$ and the effective potential assumes a runaway form for values of $\phi \gsim \phi_{crit}$. In other words, the solution transitions between the mass-lifting and ghost-like branches at $\phi_{crit}$. Note that the value taken by $\phi_{crit}$ is a function of $\hat\rho$ and the effective potential.
{\bf Bottom Left}: Plot showing the effective potential for a conformal factor as in equation~\eqref{simpleTaylor} and for different values of the parameter $k_X$ and fixed $\hat\rho$: The solid line shows the non-derivative chameleon with $k_X = 0$, dashed lines show the mass-lifting branch with $0 < k_X \hat\rho < 1$, dot-dashed lines show the mass lowering branch with $k_X < 0$ and dotted lines lie in the unstable region with $k_X \hat\rho > 1$.
{\bf Bottom Right}: Analogous plot varying $\hat\rho$ for some fixed positive $k_X$ (hence no mass-lowering branch is present here). Varying $\hat\rho$ changes the position of the minimum as well as the normalization of $V_{eff}$ and hence its curvature and mass $m^2$. $\hat\rho$ is zero for the solid line and below/above $\hat\rho_{crit}$ for dashed/dotted lines respectively.}
}
\label{figexample}
\end{figure}

In the Taylor-expanded picture laid out in the previous section, considering a separable conformal factor $A(\phi,X)$ amounts to setting $A^{(0)}(\phi) = A^{(1)}(\phi)$. An interesting feature compared with the $k_X$ case discussed above is that the normalization of the potential now also becomes a function of $\phi$. To see this explicitly we write the conformal factor as
\be
A(\phi,X) = B(\phi)C(X),
\ee
which means the effective potential satisfies
\be
V_{eff,\phi}(\phi)= \left(V_{,\phi} + \hat\rho B_{,\phi} C^{(0)}\right) \left( 1 - B C^{(1)} \hat \rho \right)^{-1},
\ee
where $C^{i}$ refers to the $i$-th component in a Taylor-series expansion of $C(X)$ in powers of $X^i$. Since at the minimum of the potential $V_{eff,\phi}(\phi_{min}) = 0$, we obtain an effective chameleon mass for oscillations around $\phi_{min}$ of
\begin{eqnarray}
m^2 &=& \frac{ V_{,\phi\phi}(\phi_{min}) -\frac{1}{2} \hat\rho C^{(0)} B_{,\phi\phi}(\phi_{min}) }{1-\frac{1}{2}k_2 B(\phi_{min}) C^{(1)} \hat\rho}.
\end{eqnarray}
Viewing the effective potential as a renormalized version of the corresponding standard (i.e. non-derivative) chameleon setup with $C = 1$, the effect of the $B(\phi_{min})$ term in the denominator can be described as making this renormalization $\phi$-dependent. The top right plot in figure~\ref{figexample} shows the effective potential for a separable conformal factor of the form $A(\phi,X) = \text{Exp}\left[k_1 \phi + k_2 X\right]$. Note that we require a positive $k_2$,  because otherwise the conformal factor diverges as $\pa_\mu \phi \to 0$, i.e. no stable, static solution is possible. Now the effective mass can be written as
\begin{eqnarray} \label{separablemass}
m^2 &=& \frac{ V_{,\phi\phi}(\phi_{min}) + \hat\rho k_1^2 e^{k_1 \phi_{min}} }{ 1-k_2 e^{k_1 \phi_{min}} \hat\rho } = m_{standard}^2 \left( 1-k_2 e^{k_1 \phi_{min}} \hat\rho \right)^{-1}.
\end{eqnarray}
The solution again separates into the three branches discussed. For positive $B(\phi)$ and when $0 < k_2 B \hat \rho < 1$ the effective potential has a well behaved minimum protected by an infinite potential barrier at $\phi_{crit}$, where $\phi_{crit}$ satisfies $k_2 B(\phi_{crit}) \hat \rho = 1$ for a given $\hat\rho$. Equation~\eqref{separablemass} then shows that $m^2$ is enhanced by the derivative-dependence of $A(\phi,X)$. For values of $\phi > \phi_{crit}$ there do exist unstable regions of parameter space. However, a particle, which starts at an initial field value $\phi_i$ for which $k_2 B(\phi_i) \hat \rho < 1$, is protected from entering the unstable region. Conversely, given a particular conformal factor that fixes $k_2$, solutions with positive $B(\phi)$ inside a source with energy density such that $\hat\rho > (k_2 B)^{-1}$ are unstable. Finally, for negative $B(\phi)$ a stable, mass-lowering solution is obtained similar to the one discussed in the previous section.
The top right plot of figure~\ref{figexample} illustrates these three branches and one can see the generation of an effective minimum and how the mass $m^2$ can be tuned by varying $k_2$ (for a fixed $\hat\rho$) at the expense of lowering the critical field value $\phi_{crit}$.

\subsection{Example III: A purely derivative conformal factor $A(X)$}

Suppose we have a purely derivative conformal factor that does not depend on the field value of $\phi$ itself. This is interesting, since it means the theory has a shift symmetry $\phi \to \phi + c$, potentially protecting it from a number of quantum corrections\footnote{This statement is of course not as strong as for a theory with an additional Galilean shift symmetry (cf.~\cite{galileon}) where an effective non-renormalization theorem can be derived~\cite{nonrenorm1,nonrenorm2}. Note, however, that even there analogous concerns enter due to the non shift-symmetric matter coupling $\phi\tilde{T}$, cf. discussions in~\cite{Hui:2009kc,ArmendarizPicon:2011ys}.} (assuming this is at most softly broken by the potential $V(\phi)$ - cf.~\cite{galinf}, where this point is discussed in an inflationary setting). As such, let us consider a conformal factor of the following form
\be
\mett = A^2(X)\met = \left(1 + A^{(1)}X + {\cal O}(X^2)\right) \met,
\ee
where $A^{(0)}$ has been appropriately normalized (i.e. there is no fundamental reason why it should be unity) and where all $A^{i}$ are now constants and therefore {\it not} functions of $\phi$. The reason to require $A^{(0)} \ne 0$ is that otherwise any coupling between gravity and matter vanishes once $\pa_\mu \phi = 0$, e.g. once $\phi$ has settled into its minimum. 
The full action can then be written as
\be \label{Sderivative}
{\cal{S}} = \int d^4 x \detg \left( \frac{M^2}{2}R + X - V(\phi)\right) +  {\cal S}_m\left( A^2(X)\met, \Psi_i \right).
\ee

Now we have already seen that a conformal factor, which does not depend on the field value $\phi$ itself, cannot create an effective potential with a minimum, if $V(\phi)$ is of the runaway form (so it does not have a mass term). But what if dark energy is in fact sourced by a very light, but not massless, cosmological scalar? Then $V(\phi)$ does already have a (very small) mass term and hence a minimum. Let us focus on a particularly simple toy model and consider a power-law potential of the form
\be
V(\phi) = \frac{1}{2}\left(\frac{\phi}{\phi_0}\right)^2,
\ee
i.e. a simple mass term for $m_\phi = \phi_0^{-1}$, which allows us to tune the mass of the field by choosing $\phi_0$. In order for this field to act as dark energy on large scales, it has to be extremely light $m_\phi = \phi_0^{-1} \lesssim H_0 \sim 10^{-33} eV$. Now the effective equation of motion for $\phi$ is
\begin{eqnarray}
V_{,\phi} = \left( 1 -\hat\rho A_{,X} \right)  \Box \phi + \hat\rho A_{,XX} \Pi - A_{,X} \pa^{\alpha} \phi \nabla_\alpha \hat\rho.
\end{eqnarray}
and we can write down an effective potential $V_{eff}(\phi) = V(\phi) \left( 1 - A^{(1)} \hat \rho \right)^{-1}$. In dense environments the effective mass of the chameleon field consequently goes as
\be
m^2 = \frac{V_{,\phi\phi}}{1 - A^{(1)}\hat\rho} = \left(\phi_0^2 - \phi_0^2 A^{(1)}\hat\rho \right)^{-1},
\ee
which can be very large subject to $0 < 1 - A^{(1)}\hat\rho \ll 1$. This outlines how a mass-lifting mechanism caused by a purely $X$-dependent conformal factor can give rise to a viable chameleon-type solution. An otherwise extremely light cosmological scalar field then acquires a large mass in dense environments. 

We emphasize that the point made here is independent of the exact form of the potential. Given any potential for a non-massless field $\phi$, i.e. a potential with a minimum at $\phi_c$, the mechanism outlined here will raise the field's mass in a density-dependent manner. We can therefore apply this purely derivative mass-lifting mechanism to any potential $V(\phi)$ for a field with a small mass $m_\phi \ne 0$, which reproduces the desired dark energy behavior on cosmological scales.

However, this approach faces a major obstacle.  The field's mass has to be enlarged by several orders of magnitude if a light cosmological scalar $\phi$ is to escape detection on solar system scales. This means $A^{(1)}\hat\rho \sim 1$ for a large range of densities $\hat\rho$, so that $A^{(1)}$ has to depend on $\hat\rho$ in order to suppress the otherwise strong dependence of $m^2$ on $\hat\rho$ and in order to prevent ghost-like instabilities from developing. We refer to section~\ref{concham2} and the previous examples in this section for a discussion of the possibility of such a density-dependent conformal factor. Also note that the mass-lowering branch with $A^{(1)} < 0$, albeit perhaps less  interesting phenomenologically, does not face this problem just as discussed in the previous examples.

\section{Radial solutions, the thin-shell effect and ``friction terms''} \label{secradial}

In this section we investigate the static, radial $\phi$-profile in and around a spherically symmetric body with uniform density $\hat \rho$ - a good approximation for the profile around the sun or earth, for example.\footnote{For a time-dependent chameleon field in the case of a radially pulsating mass, see~\cite{silvestri}.} Expressing the general equation of motion~\eqref{theeom} for such a profile one finds
\begin{eqnarray} \label{radialeom}
V_{,\phi} + A_{,\phi}\hat\rho &=& \left( 1 -\hat\rho A_{,X} \right)\left(\frac{d^2 \phi}{dr^2} + \frac{2}{r}\frac{d\phi}{dr}\right) - \hat\rho A_{,X\phi} \left(\frac{d\phi}{d r}\right)^2 \nonumber \\ &+& \hat\rho A_{,XX} \frac{d^2 \phi}{dr^2} \left( \frac{d\phi}{dr} \right)^2 - A_{,X} \pa^{\alpha} \phi \pa_\alpha \hat\rho,
\end{eqnarray}
where $\hat\rho$ and $\phi$ are functions of $r$. We now modify the approach taken in~\cite{KW2} in that we still assume 
\be
  \hat\rho(r) = \left\{ 
  \begin{array}{l l}
     \hat\rho_c & \text{for} \; r \gg R \\
     \hat\rho_\infty & \text{for} \; r \ll R. 
  \end{array} \right.
\ee
where $R$ is the radius of the spherical object, $\hat\rho_c$ is its density and $\hat\rho_{\infty}$ is the density of the surroundings.
With an eye on computing the gradient term $\pa_\alpha \hat\rho(r)$ in~\eqref{radialeom}, we model the boundary between source and surroundings by a very sharp, but smooth, transition between $\hat\rho_c$ and $\hat\rho_\infty$. The particular template we shall adopt is
\be \label{smoothrho}
\hat\rho(r) = \frac{1}{2}(\hat\rho_c-\hat\rho_\infty)(1-\text{Tanh}\left[ s(r-R) \right]) + \hat\rho_\infty,
\ee
taking $s \gg 1$, so that $\hat\rho$ effectively remains at its asymptotic values except for a sharp transition around $r = R$. A unique solution for~\eqref{radialeom} requires specifying two boundary conditions. Again following~\cite{KW2}, we take these to be $d\phi(r=0)/dr = 0$, so that the solution is non-singular at the origin, and $\phi \to \phi_\infty$ as $r \to \infty$, which ensures that the $\phi$-mediated force between two test bodies vanishes as $r \to \infty$. 

In what follows we will first recap how this setup gives rise to a thin-shell effect for the minimal conformal chameleon (i.e. the standard case). Then we discuss how radial solutions and thin-shell behavior are modified by introducing derivative conformal factors, paying special attention to the effect of ``friction terms'' in~\eqref{radialeom} (i.e. $X$ and $\Pi$ dependent terms as in ${\cal J}_2$ and ${\cal J}_3$~\eqref{Jequations}).

\subsection{The thin-shell effect for standard chameleons}

\begin{figure}[h] 
\begin{center}$
\begin{array}{cc}
\includegraphics[width=0.45\linewidth]{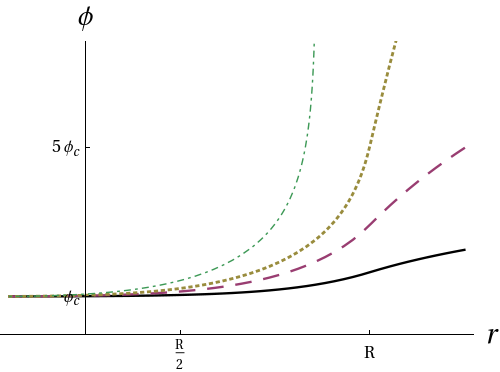} &
\includegraphics[width=0.45\linewidth]{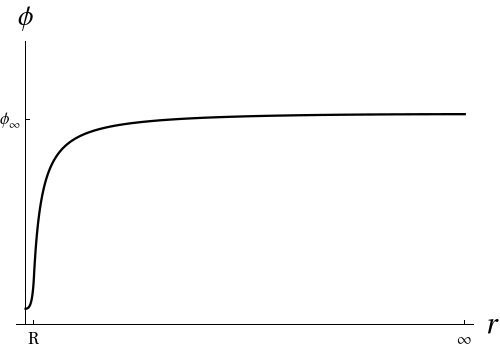}
\end{array}$
\end{center}
\caption{
\footnotesize{{\bf Left}: The radial solution for a standard chameleon with conformal factor $A(\phi) = e^{k_1 \phi}$ and different values for the parameter $k_1 = 1,1.05,1.1,1.2$ from bottom to top. The initial $\phi_i (r=0)$ remains fixed for all $k_1$ here (note that the position of the extremum $\phi_c$ also depends on $A$ and hence $k_1$).
{\bf Right}: Plot showing how a suitably chosen $\phi_i$ leads to a radial solution that asymptotes to $\phi_{\infty}$ as $r \to \infty$. $k_1 = 1 $ here.
}}
\label{figradial1}
\end{figure}

Any given physical system will come equipped with a specific $V(\phi)$, $\hat\rho$, $R$ etc. This will then allow us to compute the corresponding solution to~\eqref{radialeom}, in particular fixing the initial field-value $\phi_i$ at $r=0$. 
Alternatively one can explore the system's solutions by choosing some $\phi_i$ and finding the corresponding region in ($\hat\rho, R,...$) parameter space~\cite{KW2}.\footnote{This amounts to solving~\eqref{radialeom} as a classical mechanics problem with $\phi$ being a position- and $r$ being a time-coordinate. Note that the opposite signs for temporal and spatial dimensions in $g_{\mu\nu}$'s signature mean that, for the radial solution we are considering here, the particle whose ``position'' is described by $\phi$ ``moves'' on the inverted, and hence unstable, potential $-V_{eff}$.}   

Now for the minimal conformal chameleon, as described in section~\ref{secminimal}, the radial equation of motion is simply
\begin{eqnarray} \label{radialstandard}
V_{,\phi} + A_{,\phi}\hat\rho &=& \left(\frac{d^2 \phi}{dr^2} + \frac{2}{r}\frac{d\phi}{dr}\right).
\end{eqnarray}
Outside the source, i.e. at a radius $r > R$, the solution for $\phi$ then assumes the form~\cite{KW2}
\begin{eqnarray}
\phi(r) &\sim& \left(\frac{\Delta R}{R}\right)\frac{M_c e^{-m_{\infty}(r-R)}}{r}, \nonumber \\
\frac{\Delta R}{R} &\sim& \frac{\phi_\infty - \phi_c}{\Phi_c} \sim \frac{R-R_{roll}}{R} \ll 1, \label{thinshell}
\end{eqnarray}
where $R$ is the radius of the source, $M_c$ is its mass, $\phi_\infty$ and $m_\infty$ are the field value and mass of the field as $r \to \infty$, $\phi_c$ is the local extremum of the $\phi$-potential inside the source\footnote{Since we are effectively dealing with motion along $-V_{eff}$, $\phi_c$ is the same as $\phi_{min}$ in previous sections.} and $\Phi_c$ is the Newtonian potential at the surface of the source. We have also assumed that $\Delta R / R$, the so-called thin-shell suppression factor, is small. An intuitive picture for this setup is the following: The field is released at ``time'' $r=0$ from $\phi_i$. Inside the source the field remains stuck near $\phi_i$ with its dynamics dominated by the $2/r \cdot d\phi/dr$ friction term. Eventually, at $r = R_{roll}$, the field $\phi$ starts rolling and hence developing gradient terms. Only contributions from the region between $R_{roll}$ and $R$, i.e. the region where gradient terms develop, are felt by the exterior profile. Since $\Delta R/R \ll 1$, the chameleon force from a large massive body on a test mass is therefore thin-shell suppressed. 
This is essential in explaining why e.g. planetary orbits are not affected by a chameleonic force. In the language set out above the thin-shell regime corresponds to $\phi_i - \phi_c \ll \phi_c$. 

In contrast, the radial solution for the unsuppressed ``thick-shell'' regime, where  $\phi_i \gsim \phi_c$, is given by
\begin{eqnarray}
\phi(r) &\sim& \frac{M_c e^{-m_{\infty}(r-R)}}{r}. \label{thickshell}
\end{eqnarray}
Here the field starts significantly displaced from $\phi_c$ and consequently starts rolling almost immediately. Gradient terms develop inside the source and there is no thin-shell suppression (hence the absence of the $\Delta R / R$ factor). It is intuitively clear that the presence of further derivative terms in the general solution~\eqref{theeom} will modify this behavior, since it will affect gradient terms in $\phi$. We now move on to explore such ``friction'' terms and their effect on the thin-shell mechanism.\footnote{Further discussions of standard thin-shell screening and spherically symmetric solutions can be found e.g. in~\cite{KW1,KW2,Tamaki:2008mf,Brax:2010tj}.}

\subsection{Radial solutions for derivative chameleons}

\begin{figure}[h] 
\begin{center}$
\begin{array}{ccc}
\includegraphics[width=0.33\linewidth]{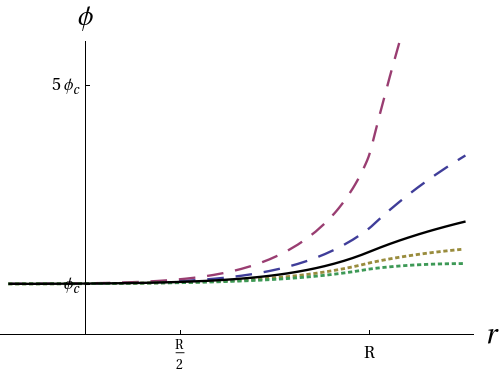} &
\includegraphics[width=0.33\linewidth]{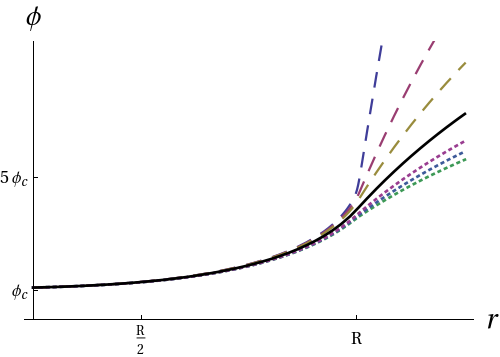} & 
\includegraphics[width=0.33\linewidth]{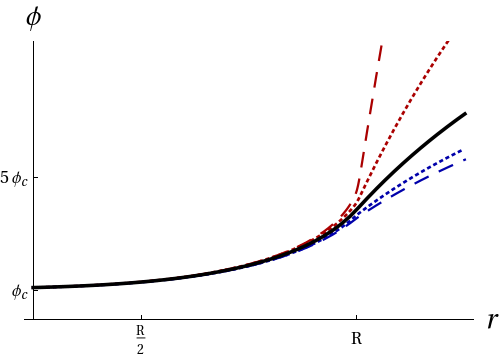}
\end{array}$
\end{center}
\caption{
\footnotesize{{\bf Left}: Plot of the radial solution for a conformal factor as in~\eqref{simpleTaylor} with $k_X = 0$ (solid line), $k_X$ increasingly positive (dashed lines - $k_X$ increases as one moves out from the solid line) and $k_X$ increasingly negative (dotted lines - $k_X$ decreases as one moves out from the solid line).
{\bf Middle}: Plot of the radial solution  for a conformal factor as in~\eqref{conformalAXP}. Again the solid line denotes $k_X = 0$. Dashed lines are for fixed positive $k_X$ and outwardly increasing $k_2$, while dotted lines are for negative $k_X$, again with outwardly increasing $k_2$. 
In order to illustrate the effect of derivative contributions more clearly the solution is shown for an initial displacement $\phi_i - \phi_c \sim {\cal O}(\phi_c / 10)$, i.e. an intermediate region between thin- and thick-shell screening. Notice that the most significant effect comes from the $\pa \hat\rho$ term at $r = R$, which produces a ``kink'' in the radial solution (actually smooth due to the density profile~\eqref{smoothrho}).
{\bf Right}: Analogous plot with the solid line describing $k_X = 0$. $k_X$ is positive above and negative below the solid line. Here dashed lines represent the solution including the friction term  that depends on $A_{,X\phi}$ (cf equation~\eqref{radialeom3}), while dotted lines ignore this contribution. The additional ``friction term'' therefore drives the solution further away from the non-derivative limit.
}}
\label{figradial2}
\end{figure}

{\bf Derivative solutions without ``friction terms''}:
How does the radial solution and thin-shell suppression change once we go beyond the simplest chameleon setups? We begin by considering a simple conformal factor of the form
\be
A(\phi, X) = e^{k_1 \phi} + k_x X,
\ee
i.e. a sum-separable $A(\phi,X) = B(\phi) + C(X)$ with purely linear dependence on $X$. This reduces~\eqref{radialeom} to
\begin{eqnarray} \label{radialeom2}
V_{,\phi} + A_{,\phi}\hat\rho &=& \left( 1 -\hat\rho A_{,X} \right)\left(\frac{d^2 \phi}{dr^2} + \frac{2}{r}\frac{d\phi}{dr}\right) - A_{,X} \pa^{\alpha} \phi \pa_\alpha \hat\rho.
\end{eqnarray}
In contrast with the standard chameleon case there are therefore two new effects. Firstly the $\left( 1 -\hat\rho A_{,X} \right)$ term, which effectively renormalizes the potential and is responsible for the mass-changing mechanism discussed in section~\ref{secconcrete}. And secondly the new gradient term in $\pa_\alpha \hat\rho$, which will only be significant very close to $r=R$. Importantly we have no $X$- or $\Pi$-dependent friction terms for a conformal factor of the form considered here.

We plot radial solutions for~\eqref{radialeom2} in the left panel of figure~\ref{figradial2} for different values of $k_x$. The behavior observed can be understood as follows. For positive $k_X$, the $\left( 1 -\hat\rho A_{,X} \right)$ term increases the curvature of the effective potential (and hence the mass of oscillations around the minimum). This means that the driving term $V_{eff,\phi}$ is enhanced and will overcome the $2/r \cdot d\phi/dr$ friction ``earlier'' in the evolution, i.e. it will reduce $R_{roll}$. In other words, in order to obtain the same thin-shell screened exterior solution one now needs to release the particle $\phi$ from even closer to the minimum value $\phi_c$, so that $\phi_{i}^{derivative} \ll \phi_i^{standard}$. The $\pa_\alpha \hat\rho$ term in fact further increases this tendency, giving an additional positive ``kick'' to the gradient of $\phi$. Note that this second effect is oblivious to the profile inside the source and as such does not modify the thin-shell condition. However, in order to reach the same boundary value $\phi_\infty$ as $r \to \infty$,  both new terms require an initial $\phi_i$ closer to $\phi_c$ than in the non-derivative case. 

For negative $k_X$ the converse is true. The curvature and associated mass of the effective potential is reduced and thin-shell screening is enhanced, i.e. $R_{roll}$ is pushed closer to $R$. This broadens the thin-shell screened parameter-space and reaching the boundary value $\phi_\infty$ as $r \to \infty$ requires an initial $\phi_i$ further away from $\phi_c$ than in the non-derivative case for negative $k_X$. 
Overall the new derivative dependent effects found here modify the parameter-space for thin-shell screened solutions, with a positive/negative $k_X$ weakening/strengthening thin-shell screening respectively. As intuitively expected from modifying the curvature of the effective potential, the mass-lifting branch is therefore associated with a suppressed thin-shell mechanism, whereas the mass-lowering branch enhances thin-shell screening. 

{\bf Derivative solutions with ``friction terms''}:
Here we finally wish to examine the effect of ``friction terms'' that depend on derivatives of $\phi$. The toy model we adopt has a conformal factor
\be \label{conformalAXP}
A(\phi,X) = e^{k_1 \phi} + k_X e^{k_2 \phi} X + {\cal O}(X^2),
\ee
where we will ignore the higher order corrections ${\cal O}(X^2)$. This means the radial solution~\eqref{radialeom} simplifies as $\Pi$-dependent contributions drop out (such contributions only come in for a conformal order with non-linear dependence on $X$).  The equation of motion consequently becomes
\begin{eqnarray} \label{radialeom3}
V_{,\phi} + A_{,\phi}\hat\rho = \left( 1 -\hat\rho A_{,X} \right)\left(\frac{d^2 \phi}{dr^2} + \frac{2}{r}\frac{d\phi}{dr}\right) - \hat\rho A_{,X\phi} \left(\frac{d\phi}{d r}\right)^2 - A_{,X} \pa^{\alpha} \phi \pa_\alpha \hat\rho.
\end{eqnarray}
In the middle and right panel of figure~\ref{figradial2} we plot solutions for constant $k_1$, but varying $k_2$ and $k_X$, essentially tuning the contribution of the $2 \hat\rho A_{,X\phi} X$ term. One can see that a large $k_2$ in combination with the presence of the $\pa_\alpha \hat\rho$ term magnifies the ``kick'' at $r \sim R$. Perhaps more interestingly, for positive $k_X$ the direct coupling of $\phi$ to its derivative in the conformal factor again counteracts the thin-shell effect, by driving the field away from near its minimum $\phi_c$. As before this results in a reduced $R_{roll}$. Also as before the converse is true for negative $k_X$. The inclusion of the $2 \hat\rho A_{,X\phi} X$ term therefore strengthens the tendencies observed above: The mass-lifting branch becomes yet more fine-tuned in order to obtain a thin-shell screened solution, while for the mass-lowering branch the parameter-space corresponding to thin-shell screening is broadened.

This concludes our brief survey of how radial solutions around a massive source are modified by the introduction of derivative terms into the conformal factor. As a generic feature, we observe that a positive $k_X$, linear dependence on X, which leads to a mass-lifting mechanism, simultaneously tightens thin-shell screening constraints, requiring a value of $\phi_i$ closer to $\phi_c$ in order to maintain a nearly constant field value inside the source. The converse is true for negative $k_X$.

\section{Disformal chameleons: A no-go theorem}\label{secdisno}

Until here we have only been considering conformal couplings to matter. Here we wish to investigate whether switching on the disformal $B^2$ term~\eqref{bim} can result in interesting modifications to the chameleon mechanism (for other disformal dark energy models we refer to~\cite{Koivisto:2008ak,disformalde}). As such, matter now couples to the full disformal matter metric 
\be
\mett = A^2(\phi,X)\met + B^2(\phi,X)\pa_\mu \phi \pa_\nu \phi.
\ee
Note that such a metric always has a well-defined inverse~\cite{bekenstein} given by
\be
\tilde{g}^{\mu\nu} = A^{-2}(\phi,X)\left( g_{\mu\nu} - B^2(\phi,X) C^{-1}(\phi,X) \pa^{\mu}\phi \pa^{\nu} \phi  \right),
\ee
subject to the condition that $A^2(\phi,X) \ne 0 \ne C(\phi,X)$, where $C = A^2(\phi,X) + B^2(\phi,X)X$.
We then have
\be \label{Sdis}
{\cal{S}} = \int d^4 x \detg \left( \frac{M^2}{2}R - \frac{1}{2}X - V(\phi)\right) +  {\cal S}_m\left(A^2(\phi,X)\met + B^2(\phi,X)\pa_\mu \phi \pa_\nu \phi, \Psi_i \right).
\ee
Just as in the conformal case we consider a non-relativistic, pressureless source. In addition we here focus on static solutions for which $\partial_0 \phi = 0$ (requiring spherical symmetry could reduce this further to $\phi = \phi(r)$). As such our assumption list now is
\begin{itemize}
\item The stress-energy tensor describes a pressureless, non-relativistic fluid ($\tilde{T}_0^0 = -\tilde{\rho}$) with all other stress-energy tensor components vanishing.
\item $\phi$'s dynamics are described by a static solution ($\partial_0 \phi = 0$).
\end{itemize}
As before we may also assume a static, uniform source in the matter frame ($\tilde{\nabla}_\alpha \tilde{T}^{\mu\nu} = 0$ inside the source), but this won't be necessary for the argument here. Instead we are solely interested in whether a chameleon mechanism can arise inside a source as a consequence of disformal contributions here at all.

The equation of motion~\eqref{phieom} then immediately gives rise to a no-go theorem. This is because all disformal ($B^2$-dependent) terms in fact involve a contraction of one of the following forms 
\be
\pa_\mu \phi \pa_\nu \phi \tmnt, \hspace{1cm} \pa_\mu \phi \tmnt, \hspace{1cm} \pa_\alpha \pa_\mu \phi \tmnt, \hspace{1cm} \pa_\mu \phi \nabla_\nu \tmnt.
\ee 
As such all disformal contributions vanish for a static solution around a non-relativistic, pressureless source, since the only non-zero component of the stress energy tensor is $\tilde{T}^{00}$, which vanishes when contracted with $\pa_0 \phi$ for a static solution. Chameleonic effects from the disformal coupling are therefore suppressed when considering e.g. solutions around the earth. We do expect disformal effects to play an important role when computing relativistic or non-static corrections to this solution, however. Nevertheless, this shows that, with the mild assumptions implemented above, the dominant contribution to the chameleon profile necessarily has to come from a conformally coupled matter metric. Disformal effects here only come in at next order in relativistic corrections and for non-static profiles. Note that we have focused on chameleonic behavior here. As comparison with the discussion of ${\cal J}_5$ and ${\cal J}_6$~\eqref{j56} in section~\ref{concham2} shows, Vainshtein-like screening can arise as a consequence of disformal couplings.\footnote{Once the metric $\mett$ is promoted to a building block for further gravitational parts of the action as in~\cite{DBIgalileon} - e.g. a cosmological-constant-like term $\sqrt{\tilde{g}} \Lambda$, an extrinsic curvature $\tilde{K}_{\mu\nu}$ and a Ricci scalar $\tilde{R}$ for $\mett$ - then a full galileon/Horndeski type solution can be obtained, making the Vainshtein screening in question more robust.}  

\section{Conclusions}\label{conc}

The main results of this paper can be summarized as follows.
\begin{itemize}
\item A derivative chameleon model described by~\eqref{Shigher}, where all matter is universally and minimally coupled to a metric $\mett = A^2(\phi,X)$, generically comes with a new mass-altering mechanism. This separates into three branches depending on the local energy density $\hat\rho$ and the conformal factor $A$: A mass-lifting (``screening'') branch, where derivative effects lead to an additional enhancement of the effective chameleon mass compared  with the non-derivative $A^2(\phi,X \to 0)$ limit. For high energy-densities above some $\hat\rho_{crit}$ this transitions into a ghost-like branch in which the effective potential becomes unstable. Thirdly there is a stable mass-lowering branch. 

\item This suggests that derivative chameleon models which are not plagued by ghost-like instabilities typically lower the effective chameleon mass. Implementing a mass-lifting mechanism while avoiding ghost-like instabilities for a range of different densities $\hat\rho$ requires some additional ``engineering'', e.g. introducing either an appropriate energy density cutoff or a density-dependent conformal factor.

\item A very light, but massive, cosmological scalar can in principle be chameleon-screened by a purely derivative-dependent conformal factor $A^2(X)$ in the mass-lifting branch, which offers added protection from quantum corrections over $A^2(\phi)$ due to the presence of the shift symmetry $\phi \to \phi + c$. 
\item The position of the minimum $\phi_{min}$ of an effective chameleon potential $V_{eff}$ cannot be affected by derivative dependent terms ($X$-dependent as well as for higher order terms). If we start with a runaway potential, e.g. $V(\phi) \sim \phi^{-n}$ for some positive $n$, this means a $\phi$-dependent conformal factor is necessary in order to produce chameleon screening. In other words, in this particular setup $A^2(\phi,X)$ can yield chameleon screening, whereas $A^2(X)$ cannot.
\item Radial solutions and the thin-shell mechanism are modified in derivative chameleon models. The region of parameter-space exhibiting thin-shell screening is reduced in the mass-lifting branch and enhanced in the mass-lowering branch of solutions.
\item Disformal contributions to the matter metric cannot source chameleon phenomenology for a static solution around a non-relativistic, pressureless source (they can source Vainshtein-screened solutions though).
\end{itemize}

To conclude, chameleon models provide a well-established framework for reconciling the presence of a light cosmological scalar that drives cosmic acceleration with small scale fifth force constraints. Here we have shown how chameleon-type setups can be generalized to derivative-dependent conformal couplings, suggesting that this naturally generates a further mass-changing mechanism. The mass-lifting branch offers the possibility to alleviate the fine-tuning involved in making a chameleon fit fifth force constraints and also comes with the exciting promise of making chameleon models more robust from a quantum perspective by endowing them with a shift symmetry. However, this branch also faces a ghost-problem which may be answered by introducing a $\hat\rho$-dependent conformal factor or a $\hat\rho$-dependent cutoff for the theory. For the mass-lowering branch no such problems exist, suggesting that a mass-suppressing mechanism is in fact a typical feature of derivative chameleon models. As such we hope that this work contributes to the enterprise of providing a wider survey of the ways in which chameleon-phenomenology might be implemented, realizing what types of challenges it has to face, and therefore putting such models on a firmer footing by showing what requirements stable chameleon models have to meet.

\vspace{1cm}
{\textbf{{\begin{large}Acknowledgements:\end{large}}}} 
I would like to thank Anupam Mazumdar for comments on an earlier draft and correspondence. I also thank Nelson Nunes and David Seery for correspondence. I am grateful to Carlo Contaldi, Ben Hoare, Jo\~ao Magueijo,  Ali Mozaffari, Silvia Nagy, Andrej Stepanchuk and James Yearsley for several helpful discussions and comments throughout this project. I am supported by the STFC.

\appendix
\section{Appendix: Covariant derivatives}

Here we present some details of the calculation for a generalized equation of motion in conformal derivative chameleon setups, expanding on the results presented in section~\ref{concham2}. We start with a metric relation
\be
\tilde{g}_{\mu\nu} = A^2\left(\phi,X\right) g_{\mu\nu}.
\ee
The covariant derivatives for the two metrics are related by
\be
\tilde{\nabla}_\alpha \omega_\beta = \nabla_\alpha \omega_\beta - \Gamma^\gamma_{\alpha\beta} \omega_\gamma.
\ee
For the Einstein frame covariant derivative $\nabla_\alpha$ acting on the matter frame stress-energy tensor, one therefore finds 
\begin{eqnarray}
\nabla_\alpha \tilde{T}^{\mu\nu} = \tilde{\nabla}_\alpha \tilde{T}^{\mu\nu} - \Gamma_{\beta\alpha}^{\mu} \tilde{T}^{\beta\nu} - \Gamma_{\beta\alpha}^{\nu} \tilde{T}^{\mu\beta}.
\end{eqnarray}
In calculating the connection $\Gamma_{\beta\alpha}^{\mu}$ explicitly, the following relations will be useful
\begin{eqnarray}
\nabla_\alpha A &=& A_{,\phi} \partial_\alpha \phi + A_{,X} \partial_\alpha X, \nonumber \\
&=&  A_{,\phi} \partial_\alpha \phi - A_{,X} \partial_\alpha \partial_\mu \phi \partial^\mu \phi. \nonumber \\
\nabla_\alpha A_{,X} &=& A_{,X\phi} \partial_\alpha \phi + A_{,XX} \partial_\alpha X, \nonumber \\
&=&  A_{,X\phi} \partial_\alpha \phi - A_{,XX} \partial_\alpha \partial_\mu \phi \partial^\mu \phi.
\end{eqnarray}
As such we can work out the connection, arriving at
\begin{eqnarray}
\Gamma^\gamma_{\alpha\beta} &=& \frac{1}{2}\tilde{g}^{\gamma \delta} \left( \nabla_\alpha \tilde{g}_{\beta\delta} + \nabla_\beta \tilde{g}_{\alpha\delta} - \nabla_\delta \tilde{g}_{\beta\alpha}   \right) \nonumber \\
&=& A^{-1} \left( 2 \delta^\gamma_{\left(\beta\right.} \nabla_{\left.\alpha\right)}A - g_{\alpha \beta} g^{\gamma \delta} \nabla_{\delta} A \right)\nonumber \\
&=& 2 A_{,\phi} A^{-1} \delta_{(\beta}^\gamma \partial_{\alpha)}\phi - 2A_{,X} A^{-1} \delta_{(\beta}^\gamma \partial_{\alpha)}\partial_{\mu} \phi \partial^{\mu}\phi \nonumber \\ 
&-& g_{\alpha\beta} g^{\gamma\delta} \partial_{\delta} \phi A^{-1} A_{,\phi} + g_{\alpha\beta} g^{\gamma\delta} \partial_{\delta} \partial_{\mu} \phi \partial^{\mu} \phi A^{-1} A_{,X}.
\end{eqnarray}

{

}

\end{document}